\documentclass[conference]{IEEEtran}
\IEEEoverridecommandlockouts

\usepackage{cite}
\usepackage{amsmath,amssymb,amsfonts}
\usepackage{algorithm, algpseudocode}
\usepackage{graphicx}
\usepackage{textcomp}
\usepackage{xcolor}
\def\BibTeX{{\rm B\kern-.05em{\sc i\kern-.025em b}\kern-.08em
    T\kern-.1667em\lower.7ex\hbox{E}\kern-.125emX}}
\usepackage{adjustbox}
\usepackage{booktabs}
\usepackage[pdfstartview=XYZ,
bookmarks=true,
colorlinks=true,
linkcolor=blue,
urlcolor=blue,
citecolor=blue,
pdftex,
bookmarks=true,
linktocpage=true, 
hyperindex=true
]{hyperref}

\usepackage{orcidlink}

\begin{document}

\title{Audio-Driven Reinforcement Learning for Head-Orientation in Naturalistic Environments\\

\thanks{This project received funding from the NWO Talent Program (VI.Veni.202.184; KH) and from the Dutch Brain Interface Initiative (DBI2; project number 024.005.022) of the NWO Gravitation research programme (YQ).\\
\\
$^1$Code is available at \url{github.com/HumanAndMachineHearing/AudioDriven_DRL_for_HeadOrientationControl}}
}

\author{\IEEEauthorblockN{Wessel Ledder \orcidlink{0009-0006-0388-208X}}
\IEEEauthorblockA{
\textit{Radboud University}\\
Nijmegen, Netherlands}
\and
\IEEEauthorblockN{Yuzhen Qin \orcidlink{0000-0003-1851-1370}}
\IEEEauthorblockA{
\textit{Donders Institute, Radboud University}\\
Nijmegen, Netherlands
}
\and
\IEEEauthorblockN{Kiki van der Heijden \orcidlink{0000-0003-4516-7907}}
\IEEEauthorblockA{
\textit{Donders Institute, Radboud University}\\
Nijmegen, Netherlands} 
\textit{Mortimer B. Zuckerman Mind Brain Behavior Institute}\\ 
Columbia University, New York, United States}

\maketitle

\begin{abstract}
Although deep reinforcement learning (DRL) approaches in audio signal processing have seen substantial progress in recent years, fully audio-driven DRL for tasks such as navigation, gaze control and head-orientation control have received little attention. Yet, such audio-driven DRL approaches are highly relevant for the development of fully autonomous audio-based agents as they can be seamlessly merged with other audio-driven (DRL) approaches such as automatic speech recognition and emotion recognition. Therefore, we propose an end-to-end, audio-driven DRL framework in which we utilize deep Q-learning to develop an autonomous agent that orients towards a talker in the acoustic environment based on stereo speech recordings. Our results show that the agent learned to perform the task in a range of naturalistic acoustic environments with varying degrees of reverberation. Quantifying the degree of generalization of the proposed DRL approach across acoustic environments revealed that policies learned by an agent trained on medium or high reverb environments generalized to low reverb environments, but policies learned by an agent trained on anechoic or low reverb environments did not generalize to medium or high reverb environments. Taken together, this study demonstrates the potential of fully audio-driven DRL for tasks such as head-orientation control. Furthermore, our findings highlight the need for training strategies that enable robust generalization across acoustic environments in order to develop real-world audio-driven DRL applications$^1$.  
\end{abstract}

\smallskip


\section{Introduction}
For natural human-robot interaction, eye-contact between a robot and a talker is an important nonverbal signal \cite{b1,b2}. Eye contact is established through gaze direction and head orientation, both of which can be controlled in robotics using deep reinforcement learning techniques (DRL, \cite{b3}). However, while DRL for robot navigation has received widespread attention \cite{b4}, few DRL approaches have been proposed for gaze control \cite{b5} and, to the best of our knowledge, none for head-orientation control. Further, most DRL approaches for robot navigation operate on visual input (e.g. \cite{b6, b7}) or on multi-modal audio-visual input (e.g. \cite{b8, b9}). Fully audio-driven approaches for robot navigation are much less prevalent \cite{b10, b11}. Thus, despite the increase in RL applications in the field of audio signal processing \cite{b12}, the potential of DRL techniques for creating audio-driven autonomous systems that learn directly from sound signals in the environment \cite{b12} remains largely unexplored for applications such as navigation, gaze control, and head-orientation. 

Here, we propose an end-to-end, audio-driven DRL approach for head-orientation control in the context of human-robot interaction. We use deep Q-learning \cite{b13} with a recurrent neural network (RNN) architecture to develop an agent that learns a policy for orienting its head towards a talker in the acoustic environment. The agent is trained directly on spatialized stereo speech recordings which contain human spatial cues \cite{b14} to improve efficiency in comparison to spectrogram-based approaches \cite{b15}. We simulate several acoustic environments (anechoic, low reverb, medium reverb and high reverb) to assess the impact of reverberation on task performance and to assess generalization across a range of naturalistic acoustic environments. Specifically, as reverberation is known to have a strong impact on human hearing (including sound localisation, \cite{b14}) as well as on machine hearing (for example, speech separation \cite{b16}), obtaining a measure of generalization across diverse naturalistic environments is crucial for future development of real-world applications \cite{b17}. 

Our results show that the proposed DRL agent can robustly learn head-orientation control from stereo speech recordings alone in a range of acoustic environments. Such fully audio-driven head-orientation control approaches can be merged seamlessly with other audio-driven (DRL) approaches, such as automatic speech recognition \cite{b12, b18} or emotion  recognition \cite{b12}, and therefore hold strong potential for fully autonomous audio-based agents. 

\begin{figure*}[t] 
\centerline{\includegraphics{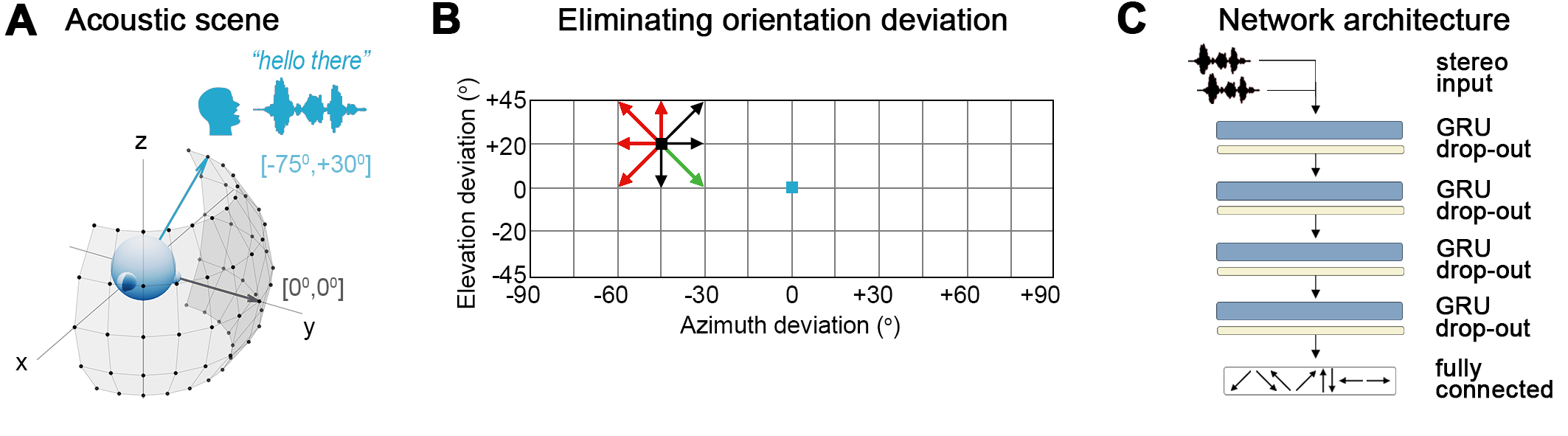}}
\caption{(A) Schematic illustration of an acoustic scene and the head orientation task. Black arrow: agent’s head orientation
at the start of the episode. Blue arrow: target head orientation. The grid reflects all possible talker locations. (B) We define the head orientation problem as a task in which the aim is to eliminate the orientation deviation between the direction of arrival of the talker's voice and the agent's head orientation. As such, we can define a 2D grid in which orientation deviation is expressed in terms of the azimuth and elevation deviation. The blue square indicates the target orientation deviation in all episodes (0\textdegree{},0\textdegree{}). The black square represents an example of an agent's orientation deviation with respect to the talker's location. Arrows indicate possible actions. Green arrow: optimal action (largest decrease in orientation deviation). Red arrows: non-optimal actions (increase in orientation deviation). Black arrows: all other actions. (C) Schematic overview of the recurrent neural network architecture used to approximate the optimal action-value function.}
\label{fig_1}
\end{figure*}

\section{Naturalistic acoustic scenes}
\label{sec:acoustic_scenes}

\subsection{Dataset}
We applied our audio-driven DRL approach to anechoic speech scenes and to naturalistic, reverberant speech scenes. Speech clips (10 s duration) were selected from the Librispeech database \cite{b19}. We selected 30 speech clips for each of 40 randomly sampled talkers to generate the training set and 20 clips of 10 different talkers to generate an independent test set. We presented the selected speech clips at 64 talker locations in the frontal hemifield: 13 azimuth locations ranging from  -90\textdegree{} to +90\textdegree{} in equidistant steps of 15\textdegree{} and 5 elevation locations (-45\textdegree{}, -20\textdegree{}, 0\textdegree{}, +20\textdegree{} and +45\textdegree{}). Talker location [0\textdegree{}, 0\textdegree{}] was excluded as this location corresponds to the agent oriented towards the talker. 
 
\subsection{Anechoic acoustic scenes}
To spatialize sound clips to a given location in an anechoic environment (that is, without reverberation), we convolved each monaural speech clip with a head related transfer function (HRTF) from the SOFA HRTF library \cite{b20}. HRTFs capture listener-specific acoustic properties and simulate human binaural listening to a source at a specific location in an anechoic environment (at 1.5 m distance). 
  
\subsection{Naturalistic acoustic scenes with reverberation}
We simulated naturalistic, reverberant rooms using binaural room impulse responses (BRIRs). First, we calculated Room Impulse Responses using PyroomAcoustics \cite{b21} for three rooms that varied in size: small (4 m x 6 m x 4 m), medium (5 m x 7 m x 4 m) and large (6 m x 8 m x 4 m). Each room was coupled with a $T_{60}$ based on the size of the room to simulate an increasing amount of reverberation with increasing room size: $T_{60}$ small room = 0.2, $T_{60}$ medium room = 0.4 and $T_{60}$ large room = 0.6. The listener was positioned centrally in the room. For each room, we generated 64 RIRs corresponding to the 64 talker locations, as well as a RIR for target location [0\textdegree{}, 0\textdegree{}]. This resulted in a total of 65 RIRs per room.  

Next, we computed BRIRs by convolving the RIRs with the HRTFs using Binaural SDM \cite{b22}. For all RIRs, we calculated a BRIR for each of the 65 talker locations and all possible head orientations of the agent. The agent's head orientation was restricted to the same range as the talker locations, that is, to 65 head orientations. In total, we generated 4,225 BRIRs per room. Finally, we convolved the  BRIRs with the monaural speech clips to create stereo clips of spatialized talkers in naturalistic, reverberant environments.  

\section{Audio-driven reinforcement learning approach}
\label{sec:reinforcementlearning}

\subsection{Problem statement}
\label{ssec: task}
Our goal is to find the shortest sequence of actions that orients the head of an agent from its starting position towards the direction of arrival of a talker's voice, such that the agent directly faces the talker (Fig.~\ref{fig_1} A). We define this problem in terms of the orientation deviation between the agent's head and the talker's location in the environment in which the agent's objective is to eliminate the orientation deviation ($OD$). Fig.~\ref{fig_1} B shows a 2-dimensional grid visualization of the task. 

$OD$ is defined as the Euclidean distance between the direction of arrival of a talker's voice (V), [${\rm azimuth}_V, {\rm elevation}_V$], and the head orientation of the agent (HOA), [${\rm azimuth}_{HOA},{\rm elevation}_{HOA}$]:

\begin{equation}
OD  = \sqrt{(\Delta {\rm azimuth})^2+(\Delta \rm {elevation})^2}
\end{equation} 

To achieve this goal, we developed a deep reinforcement learning approach which takes as input a stereo speech clip and outputs an action in 2D space. The action space consists of eight possible actions: rotating left, right, up, down, or diagonal (Fig.~\ref{fig_1} B). 

\subsection{Reinforcement learning}
\label{ssec:reinf}
We formulated the head-orientation task as a partially observable Markov Decision Process (POMDP) \cite{b23, b24} in which the agent interacts with the acoustic environment through observations, actions and rewards. The POMDP is defined as the tuple ($\mathcal{S}, \mathcal{A}, \mathcal{R}, \mathcal{P},  O, \Omega,\gamma$) in which at each time step $t \in [0,T]$, the agent receives as input observation $o_t (s_t)\in O$, which informs the agent about the state of the environment $s_t \in \mathcal S$. In our case, the observations are the segments of a speech clip. Based on the observation, the agent selects action $a_t \in \mathcal{A}$ following policy $\pi$, that is, $a_t = \pi(o_t)$ and receives reward $r(s_t) \in \rm I\!R$. The agent then transitions into the next state $s_{t+1}\in \mathcal S$ with probability $p(s_{t+1}|s_t,a_t)$ and observe $o_{t+1}\in O$ with probability $\Omega(o_{t+1}|s_{t+1})$. Note that the states to which the agent can transition, depend on its current state: $\mathcal S_{t}\subset \mathcal S$ denotes the set of possible states the agent can reach from the current state $s_{t-1}$. The aim of the reinforcement learning paradigm is to learn policy $\pi$ which maximizes the cumulative reward discounted by discount factor $\gamma$.

We utilized a semi-dense reward scheme in which the agent received a reward $r_t$ after each action $a_t$ in order to encourage the agent to learn the shortest sequence of actions to orient towards the target. To allocate this immediate reward, we compared the new orientation deviation $OD(s_{t+1})$ with the one at the previous step. Specifically,  we let
\begin{equation}
  r(s_t,a_t,s_{t+1})=\left\{
  \begin{array}{@{}ll@{}}
    +0.1, & \text{if}\  OD(s_{t+1}) = \min\limits_{s \in \mathcal S_{t+1}} OD(s), \\
    -0.2, & \text{if}\ OD(s_{t+1}) > OD(s_t), \\[4pt]
    0 & \text{otherwise}.
  \end{array}\right.
\end{equation} 

Finally, a reward $r_{target}$ of +1 was assigned upon reaching the target. If the target was not reached, the episode terminated when the agent reaches the maximum number of 20 actions. 

\begin{table}[t]
\caption{Performance of the RL agent in its training environment.}
\label{tab:perform}
\begin{center}
\begin{adjustbox}{width=\columnwidth,center}
\begin{tabular}{l|c|c|c}
\toprule
\textbf{Environment} & \textbf{Success rate (\%)} & \textbf{Chebyshev distance} & \textbf{Episode length}  \\
\toprule
Random & 27.6 & 2.8 & 17.3 \\
\textbf{Anechoic} & \textbf{100} & \textbf{0.0} & \textbf{4.3}\\
Low reverb & 66.7 & 0.5 & 12.7\\
Med reverb & 57.3 & 0.9 & 13.9\\
High reverb & 57.8 & 0.8 & 13.7\\
\bottomrule
\end{tabular}
\end{adjustbox}
\end{center}
\end{table}

\subsection{Deep Q-Learning}
\label{ssec:q-learning}
We utilized a deep Q-network (DQN) approach  \cite{b13} to train the agent. At each time step $t$, the training loop updates predicted Q-values towards target Q-values. At the start of the training loop, the agent interacts with the environment by making observation $o_t$ and taking action $a_t$ based on an $\epsilon$-greedy policy and current Q-values. The agent's experiences $e_t = (o_t,a_t,r_t,o_{t+1})$ are stored in memory replay buffer $D_t = {e_0,...,e_t}$. To ensure that equal importance is given to transitions at all orientation deviations, we structured the memory buffer by talker location with a size of 5,000 experiences per orientation deviation. This resulted in a total memory buffer size of 325,000 experiences.

The weights of the policy network $Q(o,a;\theta)$ were updated in an optimization step. To this end, we sampled experiences uniformly and at random across orientation deviations from the memory replay buffer $(o,a,r,o') \sim U(D)$. Q-values were updated at iteration $i$ based on the Huber loss \cite{b25}, which seeks to minimize:

\begin{equation}
\begin{split}
L_i(\theta_i) = & \mathbb{E}_{o,a,r,o'} \sim U(D)\Bigl[Huber\Bigl(r+\gamma \max\limits_{a'}Q(o',a';\theta^-_t) \\
&- Q(o,a;\theta_i))\Bigr)^2 \Bigr]
\end{split}
\end{equation}

We periodically adjusted target Q-values to reduce correlations between actual Q-values and target Q-values. To this end, we performed a soft update of the parameters of target network $\theta^-$ by the policy network parameters $\theta$ every $C$ steps \cite{b26}.  

\subsection{Network architecture}
We approximated the optimal action-value function with a recurrent neural network (RNN) model consisting of four GRU layers (512, 256, 128 and 64 nodes, respectively). GRU layers were followed by a drop-out layer (drop-out rate = 20 $\%$ for layer 1 - 3 and 50 $\%$ for layer 4). The final drop-out layer was followed by a fully-connected linear layer with eight outputs corresponding to the action space (Fig.~\ref{fig_1} C).   
\begin{figure}[t] 
\centerline{\includegraphics{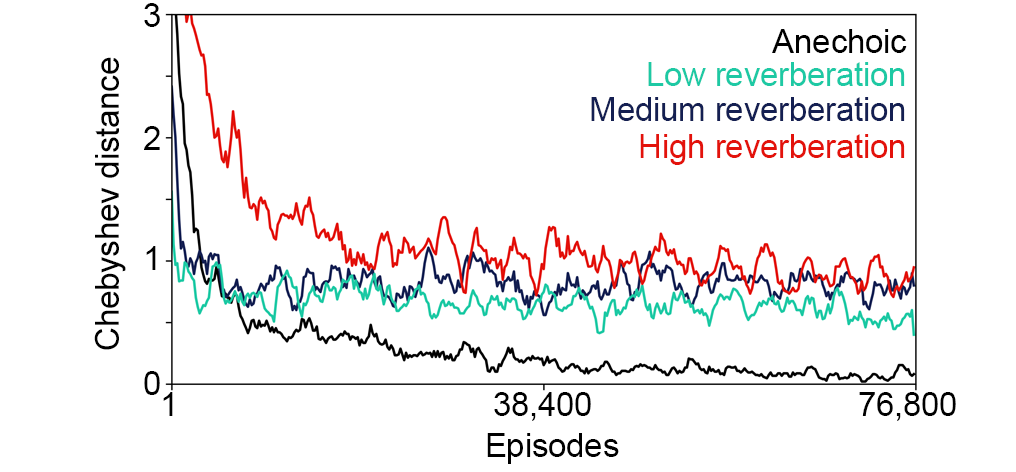}}
\caption{Temporal evolution of Chebyshev distance during training in anechoic and reverberant acoustic environments.}
\label{fig_2}
\end{figure}

\subsection{Episodes}
At the start of an episode, the selected 10 s speech clip was divided in 20 windows of 500 ms. The first window of 500 ms served as $s_0$ and was spatialized to a randomly sampled talker location in the acoustic environment. The head orientation of the agent was always [0\textdegree{},0\textdegree{}] at the start of an episode. After the agent performed action $a_t$, thereby rotating its head into a particular direction, the next state $s_{t+1}$ was generated by taking the next window of 500 ms of the speech clip and spatializing this according to the new head orientation of the agent. An episode was terminated when $OD = 0.0$ or when the maximum number of actions was reached. 

\subsection{Hyperparameters}
We developed the network with the Pytorch framework \cite{b27}, using AdamW Optimizer \cite{b28}, a learning rate of 0.001 and a batch size of 1024. We set discount factor $\gamma$ to 0.8 and reduced $\epsilon$ linearly from 0.2 to 0.0 over the course of 30,000 episodes. The agent was trained for a total of 76,800 episodes. Target network update parameters were $C = 1$ and $\tau = 0.00005$.

\subsection{Evaluation procedure}
We evaluated trained agents on 192 episodes which were generated with an independent set of talkers. We report success rate (percentage of episodes in which the agent successfully oriented towards the talker location), remaining Chebyshev distance (that is, the minimum number of actions still needed to reduce the orientation deviation to zero), and episode length (that is, the average number of actions). The performance of trained agents was compared to the performance of an agent behaving at random.  

\section{Experiments and results}
\label{sec:experiments}

\subsection{Audio-driven head-orientation within training environment}
\label{subsec: natscenes}
The evolution of the Chebyshev distance metric shows that the audio-driven DQN agents learned to orient towards a talker over the course of training, both in anechoic and in naturalistic environments with reverberation (Fig.~\ref{fig_2}). Evaluating the trained agents on the independent test episodes demonstrated that the RL agent trained on anechoic environments was highly successful at the task: the agent oriented towards the talker location on 100 \% of the episodes (average remaining Chebyshev distance = 0.0), taking an average of 4.3 actions to reach the target (Table.~\ref{tab:perform}, average shortest sequences = 3.4). 

Performance was notably lower when the audio-driven DQN agent was trained on naturalistic environments with reverberation: In the presence of low, medium or high reverberation, the agent's success rate dropped to 66.7 \%, 57.3 \%, and 57.8 \%, respectively (Table.~\ref{tab:perform}). Nevertheless, for episodes in which the agent did not reach the target orientation, the final orientation deviation was small: Chebyshev distance was $<$ 1 for all naturalistic acoustic environments (Table.~\ref{tab:perform}). Moreover, our audio-driven DQN approach still substantially outperformed an agent performing at random (success rate = 27.6 \%). 

\begin{table}[t]
\centering
\caption{Generalization across acoustic environments}
\begin{tabular}{l|l|c|c|c}
\toprule
\textbf{Training} & \textbf{Testing} & \textbf{Success} & \textbf{Chebyshev} & \textbf{Episode} \\
\textbf{environment} & \textbf{environment} & \textbf{rate \%} & \textbf{distance} & \textbf{length} \\
\toprule
Anechoic & Low reverb & 62.5 & 0.8 & 13.6  \\
 & Med reverb & 46.4 & 1.9 & 15.0 \\
 & High reverb & 29.7 & 2.7 & 16.4 \\
\midrule
Low reverb & Anechoic & 71.4 & 0.5 & 12.8  \\
 & Med reverb & 35.9 & 1.8 & 16.0 \\
 & High reverb & 25.0 & 3.0 & 17.1 \\
\midrule
Med reverb & Anechoic & 56.2 & 0.9 & 14.3  \\
 & Low reverb & 60.9 & 0.7 & 13.5 \\
 & High reverb & 41.7 & 1.6 & 15.5 \\
\midrule
High reverb & Anechoic & 32.8 & 1.3 & 16.8  \\
 & Low reverb & 48.4 & 1.0 & 14.5 \\
 & Med reverb & 58.3 & 0.7 & 14.0 \\ 
\bottomrule
\end{tabular}
\label{tab:generalization}
\end{table}

\subsection{Generalization across acoustic scenes}
\label{subsec: generalization}
\begin{figure}[t] 
\centerline{\includegraphics{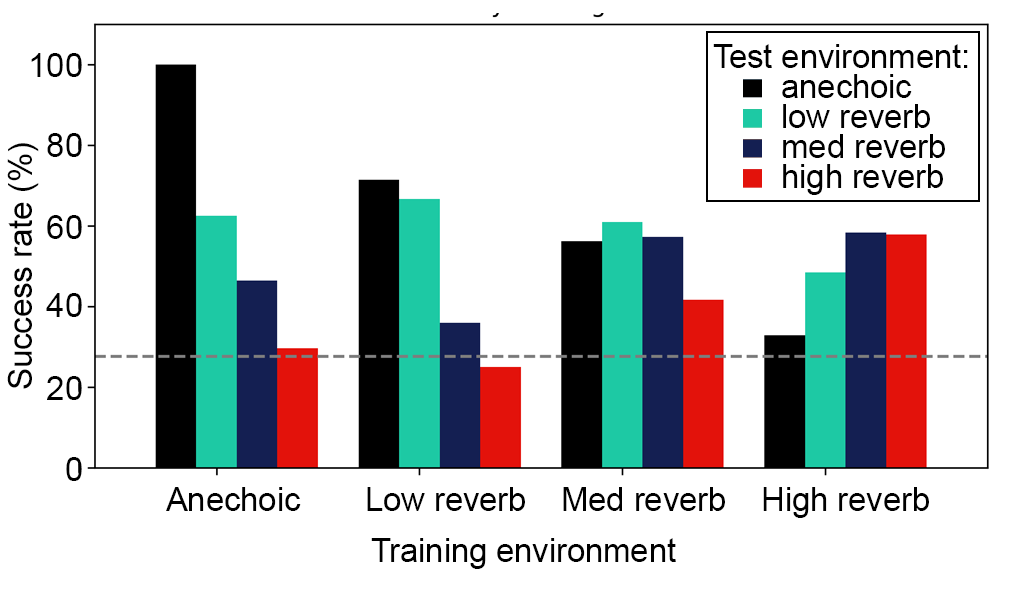}}
\caption{Generalization of trained agent to new  acoustic environments. Bars reflect performance of the trained agent on evaluation episodes for each of the acoustic environments. Dashed line indicates success rate of random agent.}
\label{fig_4}
\end{figure}

To evaluate to what degree the audio-driven DQN agents generalize across acoustic environments, we evaluated performance of the trained agents in environments with reverberation properties that deviated from the training environment (i.e., in out-of-distribution settings). Table.~\ref{tab:generalization} and Fig.~\ref{fig_4} show that the performance of the agent trained on anechoic environments was substantially lower in naturalistic environments with reverberation. Specifically, there was an inverse relationship between reverberation strength in the testing environment and performance of the agent trained on anechoic environments. 

Overall, the results indicate that agents generalize to acoustic environments with less reverberation than the environment they were trained on, but not to environments with more reverberation. However, the agent trained on high reverb environments did not perform well in anechoic environments (Table.~\ref{tab:generalization}, Fig.~\ref{fig_4}). 

\begin{figure}[t] 
\centerline{\includegraphics{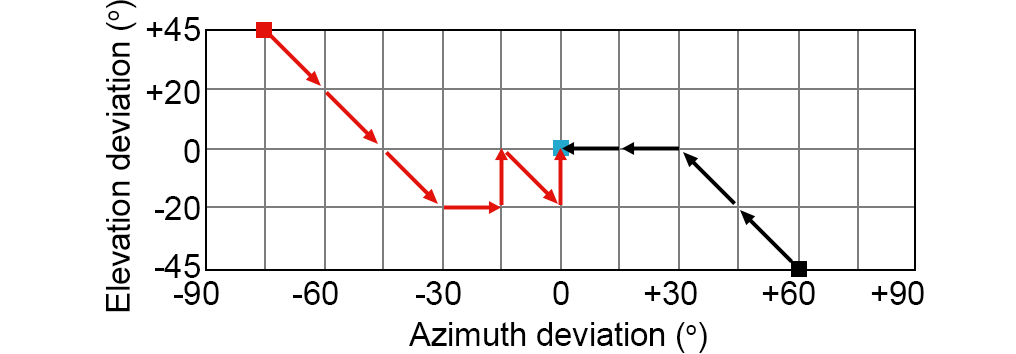}}
\caption{Example of two trajectories of the trained DQN agents. Black arrows: trajectory of an agent trained on anechoic environments. Red arrows: trajectory of an agent trained on a high reverb environment.}
\label{fig_3}
\end{figure}

\section{Conclusion}
We introduced a novel audio-driven DRL approach for head-orientation control in the context of human-robot interaction. The autonomous agent achieved near perfect performance in anechoic environments. While the presence of reverberation in naturalistic acoustic environments affected the agent's performance, the agent substantially outperformed an agent behaving at random in such reverberant environments. Further, our results show that trained agents generalize to environments with less reverberation than the training environment. In summary, we show that audio-driven DRL is a promising approach to create autonomous systems that learn directly from sound signals in the environment.

\end{document}